# An Intelligent Approach to Software Cost Prediction

Xishi Huang, Danny Ho[1], Luiz F. Capretz, Jing Ren

Dept. of ECE, University of Western Ontario, London, Ontario, N6G 1H1, Canada

[1]Toronto Design Center, Motorola Canada Ltd., Markham, Ontario, L6G 1B3, Canada

{xhuang22, lcapretz, jren2}@uwo.ca, Danny.Ho@motorola.com

**Abstract**

Good software cost prediction is important for effective project management such as budgeting, project planning and control. In this paper, we present an intelligent approach to software cost prediction. By integrating the neuro-fuzzy technique with the well-accepted COCOMO model, our approach can make the best use of both expert knowledge and historical project data. Its major advantages include learning ability, good interpretability, and robustness to imprecise and uncertain inputs. The validation using industry project data shows that the model greatly improves prediction accuracy in comparison with the COCOMO model.

## 1    Introduction

Software development is notorious for going over time and budget and the development cost is difficult to estimate beforehand. This problem lies in the fact that software development is a complex process due to the number of factors involved, including the human factor, the complexity of the product that is developed, the variety of development platforms, and the difficulty of managing large projects. As software development has become an essential investment for many organizations, accurate software cost prediction models are needed to effectively predict, monitor, control and assess software development. Since prediction accuracy is largely affected by modeling accuracy, finding good models for software prediction is now one of the most important objectives of the software engineering community.

Several software cost prediction models [1]-[5] have been developed over the last decades. COCOMO [1], [2] is arguably the most popular and widely used software prediction model. Since the cost prediction problem is complex and has features such as complex nonlinear relationships, imprecise and uncertain measurement of software metrics, no model has been proven to provide the perfect solution so far. Neuro-fuzzy technique is to some extent superior to other approaches since this approach inherently possesses learning/adaptive ability and the capability of handling imprecise information. Although neuro-fuzzy technique is scarcely investigated and far less mature than other techniques in this field, it emerges as a very promising approach that is worth much more attention. In this paper, by integrating the neuro-fuzzy technique with the well-accepted COCOMO model, we put forward an intelligent approach for software cost prediction.

## 2    Intelligent Approach to Software Cost Prediction

Our intelligent approach is based on the neuro-fuzzy technique [6]-[8] and the COCOMO model. The structure of the developed intelligent model is shown in Figure 1. The inputs of this model are the software size and ratings of 22 cost drivers including five scale factors ($SFR_i$) and seventeen effort multipliers ($EMR_i$).   The output is the software development effort prediction. Ratings of cost drivers can be given by continuous numerical values or linguistic terms such as "Low", "Nominal" and "High". The parameters in this model are calibrated by learning from industry project data.

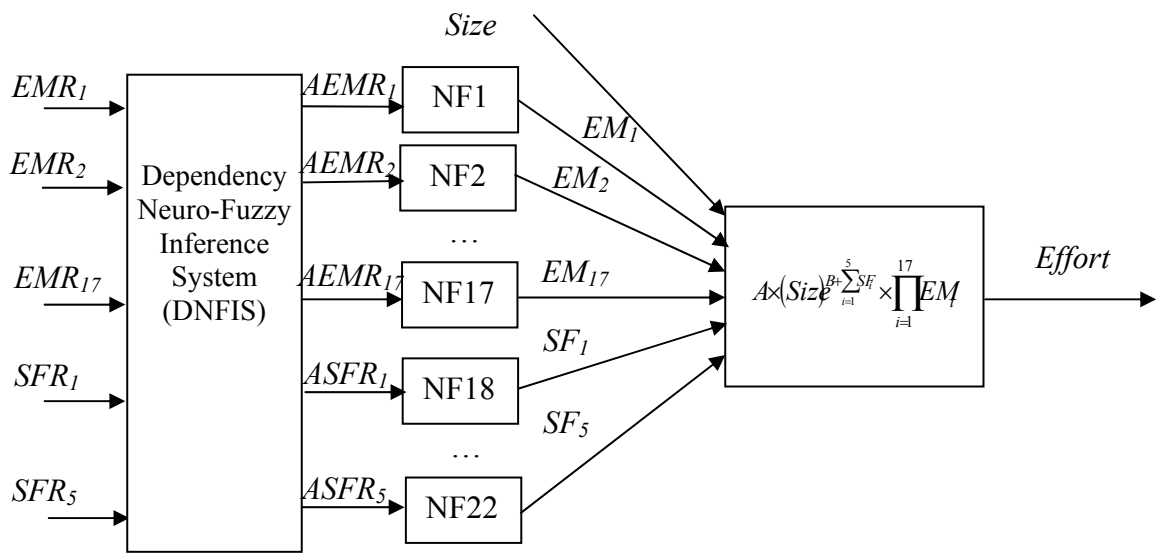

Figure 1. Intelligent Model for Software Effort Prediction

There are 3 major components in the intelligent model:

- Dependency Neuro-Fuzzy Inference System (DNFIS): This component handles the dependencies among cost drivers. The inputs are the ratings of the cost drivers, and its output is the adjusted ratings of cost drivers.

- Neuro-fuzzy subsystem (NF$_i$): The input is the adjusted rating of the *i*-th cost driver, the output is the multiplier value of the *i*-th cost driver that is used for the input of COCOMO II Model.

- COCOMO II Model: $effort = A \times (Size)^{B + \sum_{i=1}^{5} SF_i} \times \prod_{i=1}^{17} EM_i$

## 2.1 Dependency Neuro-Fuzzy Inference System (DNFIS)

Because software development is a very complex process, there are many factors contributing to development cost and there exists complex interaction between these factors. In the COCOMO model, it is assumed that the contributions of cost drivers to software cost are independent, but experts diverge from this assumption. For example, contribution of analyst capability (ACAP) and product complexity (CPLX) are not independent, which is referred to as dependency between ACAP and CPLX for short in this paper.

In COCOMO II.2000, if ACAP is rated as Very High, the software cost should be reduced to 71% of that with a Nominal rating. This implies that rating VH of ACAP will reduce software cost by 29% no matter how the product complexity is. If this is true for the Nominal CPLX, we believe that productivity cannot be increased so much when the same team develops a software product with Extra High complexity. That is, if CPLX is Extra High, ACAP should be rated lower than in the case of CPLX is Nominal. In other words, a team with VH ACAP needs more effort to develop a product with the Extra High complexity; it is likely that a team with VL ACAP will develop the same product with much higher cost. As a result, properly handling of the dependencies among cost drivers becomes essential to improve prediction accuracy.

Unfortunately, this is not an easy task in most cases; dependency is inherently a very complex problem as we have to consider the effect of any possible combinations. It is common sense that when it comes to a very complicated situation, experts usually have good understandings of the dependency. Therefore better utilization of expert knowledge is the key to handle this dependency problem. We propose the use of a dependency neuro-fuzzy inference system (DNFIS) that encodes expert knowledge into fuzzy if-then rules. For example, the following fuzzy rule: "***IF CPLX is Extra High, THEN ACAP is lowered***" is added into DNFIS for the above case. DNFIS can effectively resolve the dependencies among the cost drivers. The inputs of DNFIS are the ratings of the cost drivers ($CDR_i$s), and its output is the adjusted ratings of cost drivers ($ACDR_i$s).

### 2.2   Neuro-Fuzzy Subsystem $NF_i$

We adopt the Adaptive Neuro-Fuzzy Inference System (ANFIS) [8] for each neuro-fuzzy subsystem $NF_i$. We denote $CD_i = SF_i$, $i = 1,2,\cdots,5$ and $CD_{i+5} = EM_i$, $i = 1,2,\cdots,17$. The input of $NF_i$ is the adjusted rating value $ACDR_i$ of the i-th cost driver $CD_i$, and the output is the corresponding numerical/multiplier value $CD_i$. Figure 2 shows the structure of subsystem $NF_i$, which is functionally equivalent to a Takaki and Sugeno's [11] type of fuzzy system with the following rules:

Fuzzy Rule *1*: If $ACDR_i$ is $A_{i1}$ (Very Low), then $CD_i=CD_{i1}$,

Fuzzy Rule 2: If $ACDR_i$ is $A_{i2}$ (Low), then $CD_i=CD_{i2}$,

Fuzzy Rule *3*: If $ACDR_i$ is $A_{i3}$ (Nominal), then $CD_i=CD_{i3}$,

Fuzzy Rule *4*: If $ACDR_i$ is $A_{i4}$ (High), then $CD_i=CD_{i4}$,

Fuzzy Rule *5*: If $ACDR_i$ is $A_{i5}$ (Very High), then $CD_i=CD_{i5}$,

Fuzzy Rule *6*: If $ACDR_i$ is $A_{i6}$ (Extra High), then $CD_i=CD_{i6}$,

where the fuzzy set $A_{ik}$ corresponds to a rating level ranging from "Very Low" to "Extra High" for the i-th cost driver, and $CD_{ik}$ is the corresponding parameter value of the k-th rating level.

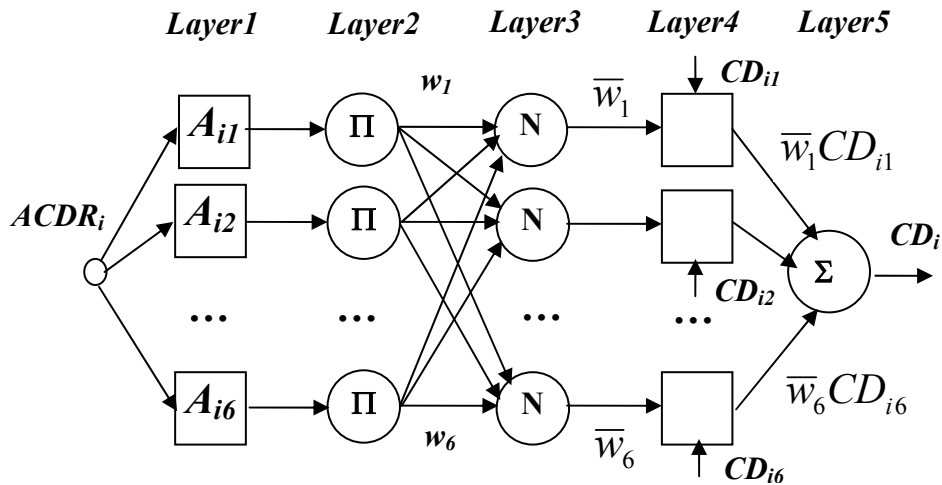

Figure 2. The Structure of Subsystem $NF_i$

Now we describe the functions of the subsystem $NF_i$ layer by layer.

_Layer 1_: This layer is used to calculate the membership values for each rule. The activation function of node in this layer is defined as the corresponding membership function.

$$O_k^1 = \mu_{ik}(ACDR_i)$$

where $ACDR_i$ is the adjusted rating value of the i-th cost driver, and $A_{ik}$ is a fuzzy set associated with the k-th rating level such as Low, High. $\mu_{ik}(ACDR_i)$ is the membership function of $A_{ik}$.

_Layer 2_: This layer calculates the firing strength for each rule. For each node, the inputs are the membership values in the premise of the fuzzy rule. The output is the product of all input membership values, which is called the firing strength of the corresponding fuzzy rule. Because there is only one condition in the premise of each fuzzy rule of $NFi$, the firing strength is the same as the membership value obtained from Layer 1.

$$w_k = O_k^1 .$$

_Layer 3_: This layer is used to normalize the firing strength for each fuzzy rule. The output of the k-th node is called the normalized firing strength, which is defined as:

$$\overline{w}_k = \frac{w_k}{\sum_{j=1}^{6} w_j}, \qquad k = 1,2,\cdots,6$$

_Layer 4_: In this layer, the reasoning result of a rule is calculated as follows:

$$O_k^4 = \overline{w}_k CD_{ik}$$

Parameters $\{CD_{ik}\}$ are calibrated through learning from numerical project data.

_Layer 5_: This layer sums up all the reasoning results of fuzzy rules we get from Layer 4, i.e.,

$$O_k^5 = \sum_k O_k^4 .$$

In summary, the overall output of neuro-fuzzy subsystem $NF_i$ is

$$CD_i = \sum_k \overline{w}_k CD_{ik} = \sum_k \frac{\mu_{ik}(ACDR_i)}{\sum_j \mu_{ij}(ACDR_i)} CD_{ik} .$$

2.3  Learning Algorithms

The developed intelligent model can be described by the following formulas.

For the dependency neuro-fuzzy inference system (DNFIS), the adjusted rating $ACDR_i$ of the i-th cost driver can be calculated by:

$$ACDR_i = f_{DNFi}(CDR_1, CDR_2, \cdots, CDR_{22}; P_{DNFi}), \quad i = 1,2,\cdots,22 \qquad (1)$$

where $P_{DNFi}$ is the parameter vector calibrated by the learning process.

For the neuro-fuzzy subsystem NFi, we use the following formula to calculate the numerical/multiplier value $CD_i$ of cost driver $i$,

$$CD_i = f_{NFi}(ACDR_i; P_{NFi}) = \sum_k \frac{\mu_{ik}(ACDR_i)}{\sum_j \mu_{ij}(ACDR_i)} CD_{ik}, \quad i = 1,2,\cdots,22 \qquad (2)$$

where $P_{NFi} = [CD_{i1}, CD_{i2}, \cdots, CD_{i6}]$ is the parameter vector calibrated by the learning process.

For the COCOMO II post architecture model, we predict the software development effort by:

$$\begin{aligned}
Effort &= f_{CM}(CD_1, CD_2, ..., CD_{22}, Size) \\
&= A \times (Size)^{B+0.01 \times \sum_{i=1}^{5} SF_i} \times \prod_{i=1}^{17} EM_i \\
&= A \times (Size)^{B+0.01 \times \sum_{i=1}^{5} CD_i} \times \prod_{i=6}^{22} CD_i
\end{aligned} \qquad (3)$$

From (1) to (3), we can rewrite our intelligent model as follows:

$$Effort = f_{NF}(X, P) \qquad (4)$$

where

$X = [EMR_1 \ EMR_2 \ ... \ EMR_{17} \ SFR_1 \ SFR_2 \ ... \ SFR_5 \ Size]$ denotes the input vector of the intelligent model, $P = [P_{NF1} \ ... \ P_{NF22}; P_{DNF1} \ ... \ P_{DNF22}]$ stands for the adjustable parameters in the model.

Now let us derive the learning algorithm. Given $NN$ project data points $(X_n, E_{dn})$, $n = 1,2,\cdots,NN$ ($X_n$ is the software size and the rating values of cost drivers for Project $n$, $E_{dn}$ is the actual effort for Project $n$), the learning problem of parameters $P$ can be formulated as the following optimizing problem:

$$E = \sum_{n=1}^{NN} \frac{1}{2} w_n \left( \frac{E_n - E_{dn}}{E_{dn}} \right)^2 \qquad (5)$$

subject to the following monotonic constraints [12]:

$$CD_{i1} \leq CD_{i2} \leq CD_{i3} \leq CD_{i4} \leq CD_{i5} \leq CD_{i6}, i \in I_{INC}(CD) \qquad (6)$$

$$CD_{i1} \geq CD_{i2} \geq CD_{i3} \geq CD_{i4} \geq CD_{i5} \geq CD_{i6}, i \in I_{DEC}(CD) \qquad (7)$$

where

- $w_n$ is the weight of project $n$,

- $E_n = Effort_n = f_{NF}(X, P)\big|_{X=X_n}$ is the predicted effort by the model,

- $I_{INC}(CD)$ is the set of increasing cost drivers whose higher rating values correspond to higher development effort,

- $I_{DEC}(CD)$ is the set of decreasing cost drivers whose higher rating values correspond to lower development effort.

The learning algorithm is as follows:

$$P^{l+1} = P^l - \alpha \frac{\partial E}{\partial P} \qquad (8)$$

where $\alpha > 0$ is the learning rate, $l$ is the current iteration index,

$$\frac{\partial E}{\partial P} = \sum_{n=1}^{NN} \frac{w_n}{E_{dn}^2} (E_n - E_{dn}) \frac{\partial E_n}{\partial P} \qquad (9)$$

$$\frac{\partial E_n}{\partial P_{NFi}} = \frac{\partial f_{CM}}{\partial CD_i} \cdot \frac{\partial f_{NFi}}{\partial P_{NFi}} \qquad (10)$$

$$\frac{\partial E_n}{\partial P_{DNFi}} = \frac{\partial f_{CM}}{\partial CD_i} \cdot \frac{\partial f_{NFi}}{\partial ACDR_i} \cdot \frac{\partial f_{DNFi}}{\partial P_{DNFi}} \qquad (11)$$

## 2.4 Major Features of the Intelligent Model

- **Learning ability**: An inherent feature of the neuro-fuzzy approach is the learning/adaptation capability. By training the proposed model with industry project data, the model can approximate any highly complex nonlinear relationships between software development effort and cost drivers to any accuracy.

- **Robust to imprecise and uncertain inputs**: By allowing for continuous rating values, the proposed model is insensitive to the imprecision and uncertainty in inputs such as ratings of the cost drivers. For example, for two similar projects studied in the experiments, COCOMOII produces estimates of 203 staff-months and 2886 staff-months, respectively. These are divergent results to similar projects. On the other hand, our approach gives around 809 staff-months for both projects [12], which makes more sense, and should be close to the reality.

- ***Good Interpretability***: Although the neural network approach is good at modeling complex relationships, it has some inherent shortcomings: neither is it easy to understand nor is it easy to explain its decision process [7]. This shortcoming greatly impedes the approach from gaining wide acceptance in the software engineering community. The intelligent model, by introducing fuzzy rules into the model, can effectively overcome the above problem. These fuzzy rules mimic the human reasoning in making decisions, and therefore the decision-making process in our model is clear to users and can be interpreted and validated by experts. Consequently, our model can be easily accepted for project management.

- ***Knowledge integration***: In our model, we integrate expert knowledge with numerical project data by using fuzzy rules. By making the best use of information from different sources in the decision-making process, the model can achieve more accurate and reasonable cost prediction than conventional approaches. For example, we can integrate monotonic constraints, which reflect expert knowledge of cost drivers, into our model to guarantee that the calibration results are reasonable.

- ***Local learning***: In our model, the learning parameters are decoupled. This feature allows the model to learn just some of the parameters each time; in other words, the model can accumulate knowledge locally.

## 3    Experiment Results

We use industry project data to validate our model. There is a total of 69 project data available, including 6 project data from the industry [9], [10] and 63 project data from the original COCOMO'81 database [2]. We were not able to use the COCOMOII database due to absence of individual cost driver ratings for the project data. Because most of the project data are compatible only with the intermediate COCOMO'81 model, we use the COCOMO'81 model for our validation. The COCOMO'81 model contains only 15 effort multipliers, yet our model is completely compatible with the COCOMO II model. The experiment results are shown in Table 1. PRED stands for the prediction at level $p$, i.e. $PRED(p) = k/N$, where $N$ is the total number of projects, $k$ is the number of projects with absolute relative error of $p$. If we consider projects with relative error within 20% of actual efforts, we notice that cost estimation accuracy, using the intelligent model, has improved by about 18% when compared with the COCOMO'81 model.

Table 1. Effort prediction for all 69 project data points

| PRED | COCOMO81 Model | | Intelligent Model | | Improvement |
|------|------------|----------|------------|----------|-------------|
|      | # Projects | Accuracy | # Projects | Accuracy |             |
| 20%  | 49         | 71%      | 62         | 89%      | 18%         |
| 30%  | 56         | 81%      | 64         | 92%      | 11%         |
| 50%  | 65         | 94%      | 67         | 97%      | 3%          |
| 100% | 69         | 100%     | 69         | 100%     | 0%          |

## 4    Conclusion

In this paper, we present an intelligent approach to software cost prediction. In this approach, we employ a dependency neuro-fuzzy inference system (DNFIS) to effectively handle the dependencies among cost drivers, and twenty-two neuro-fuzzy subsystems to calibrate the parameters in the COCOMO model. By integrating the advantages of three complementary techniques (neural networks, fuzzy logic and COCOMO), this intelligent approach has learning ability, knowledge integration capability, good interpretability, and robustness to imprecise and uncertain inputs. The validation using industry project data confirms that the model greatly improves prediction accuracy in comparison with the COCOMO model.

Finally, because the neuro-fuzzy technique allows the integration of expert knowledge and numerical project data, it can be a powerful tool to tackle important problems in software engineering such as cost and quality prediction. Therefore, a promising line of future work is to extend the neuro-fuzzy approach to other cost and quality prediction models and tools such as COQUALMO, SLIM, SPR knowledgePLAN, and CA-Estimacs, among others.